\documentclass{article}
\usepackage[margin=3cm]{geometry}

\usepackage{authblk}

\usepackage{amsmath,amssymb,amsfonts}
\usepackage{mathrsfs} 
\usepackage{mathtools}
\usepackage{physics}

\usepackage[nocompress]{cite}

\renewcommand{\dd}{\mathrm{d}}

\numberwithin{equation}{section}

\usepackage{hyperref}
\hypersetup{colorlinks=true,allcolors=blue,urlcolor=cyan}

\title{Azimuthal geodesics in closed FLRW cosmologies}

\author[1,2]{Christian G B\"ohmer\footnote{Email: c.boehmer@ucl.ac.uk}}
\author[1]{Antonio d'Alfonso del Sordo\footnote{Email: a.dalfonsodelsordo@ucl.ac.uk}}
\author[1]{Betti Hartmann\footnote{Email: b.hartmann@ucl.ac.uk}}
\affil[1]{Department of Mathematics, University College London, \authorcr Gower Street, London WC1E 6BT, UK\medskip}
\affil[2]{Astrophysics Research Centre, School of Mathematics, \authorcr Statistics and Computer Science, University of KwaZulu-Natal, \authorcr Private Bag X54001, Durban 4000, South Africa\medskip}

\date{\today}

\begin{document}

\maketitle

\begin{abstract}
Modern cosmology is closely linked to our understanding of radial null geodesics as these model the propagation of light signals through an expanding universe. Azimuthal geodesics, on the other hand, are perhaps best known for their relevance within closed cosmological models. Such models typically have a finite lifetime: the universe expands up to a maximum size after which it recollapses during the so-called {\it big crunch}. An azimuthal geodesic starting at the beginning of the universe will travel a finite angular distance during the expansion and recollapse. It is well-known that this angle is $2\pi$ for a matter-dominated universe and $\pi$ for a radiation-dominated solution. Here we derive the simple formula
\begin{equation*}
    \Delta \varphi = \frac{2\pi}{1+3w}
\end{equation*}
for an arbitrary linear equation of state parameter $w$. To the best of our knowledge this result has not been reported elsewhere and fills a small gap in the literature.
\end{abstract}

\section{Introduction}

In the second decade of the last century, the theory of General Relativity (GR) not only revolutionised our understanding of space and time, but also
led to the advent of Modern Cosmology. GR, which  has been tested and verified in many observations since its introduction by Einstein in 1915~\cite{Einstein:1916vd}, assumes that spacetime is dynamical and that its curvature is caused by its energy-momentum content. Gravity is a consequence of spacetime curvature and hence dictates the motion of energy/matter. This is often summarised in Wheeler's famous statement, ``spacetime tells matter how to move; matter tells spacetime how to curve''~\cite{Misner:1973prb}. The equation that connects the geometry of spacetime curvature to its energy-momentum content is the Einstein equation. In the most general setting, the Einstein equation leads to a set of 10 independent non-linear coupled partial differential equations, known as the Einstein field equations. To solve these equations is a formidable task which is, in general, only possible numerically. However, imposing spacetime symmetries can lead to a remarkable simplification of the equations. This is exactly what happens when studying cosmological models and is based on the observational fact that the universe is spatially homogeneous and isotropic on large scales, an idea known as the \emph{cosmological principle}. In simpler terms, it states that we, as observers on Earth, have no preferred position in the universe and that the universe looks the same in all directions. The distribution of galaxies in the universe is an indicator of homogeneity~\cite{Yadav:2005vv}, whilst a crucial piece of evidence supporting isotropy is the isotropic nature of the Cosmic Microwave background (CMB)~\cite{Planck:2019evm}, which is the remnant radiation from the Big Bang. 

The mathematical models describing an expanding, spatially homogeneous, and isotropic universe were studied independently by Friedmann~\cite{friedman1922krummung}, Lema\^{i}tre~\cite{Lemaitre:1927zz}, Robertson~\cite{Robertson:1935zz}, and Walker~\cite{Walker:1937qxv}, often referred to as FLRW. Mathematically, the cosmological principle leads to a spacetime which is a 4-dimensional Lorentzian manifold with 3-dimensional spatial sections that are {\it maximally symmetric} (and hence isotropic and homogeneous); topologically, the spacetime has the structure $\mathbb{R} \times {\cal M}_3$. The 3-dimensional
manifold ${\cal M}_3$ can have constant positive, vanishing or negative curvature. This is encoded by the curvature parameter $k$ which can have the following values
\begin{itemize}
    \item $k=-1$, constant negative curvature, hyperbolic space (``open'' universe),
    \item $k=0$, no curvature, Euclidean space (``flat'' universe),
    \item $k=1$, constant positive curvature, 3-sphere (``closed'' universe).
\end{itemize}

In the context of GR, a \emph{geodesic} is the path followed by a free-falling particle or light ray through spacetime. Since FLRW is isotropic, we can describe it using spherical spatial coordinates. In spherical coordinates, the azimuthal angle measures the angular position in the horizontal plane, and hence, for an azimuthal geodesic, the motion occurs in the azimuthal direction. Azimuthal geodesics are particularly helpful when describing motion in rotating or spherical spacetimes, such as black holes spacetimes. Here the azimuthal geodesics would represent the paths that particles or light rays follow as they orbit the black hole. 

Many textbooks, for example~\cite{problembook,Wald:1984rg,Boehmer:2016ome}, set the computation of the angular distance travelled along an azimuthal geodesic during the evolution of a closed universe as an exercise. However, these exercises are restricted to two special cases: the matter dominated universe and the radiation dominated universe, since in these cases one can find simple closed form solutions to the field equations. It so happens that the angles computed in these two settings differ by a factor of two. In the following we will show that this angle can be found for all cosmological models with linear equation of state.

\section{Cosmological field equations}

The FLRW line element is given by
\begin{equation}
    \dd s^2=-\dd t^2+a^2(t)\left[\frac{\dd r^2}{1-kr^2}+r^2\dd\theta^2+r^2\sin^2\negmedspace\theta \dd\varphi^2\right]\,,
    \label{FLRW}
\end{equation}
where $a(t)$ is the \emph{scale factor}, which quantifies the expansion of the universe. We further assume that the energy/matter content of the universe is described well by a perfect fluid with energy density $\rho(t)$ and pressure $p(t)$. The non-vanishing components of the Einstein equation then read 
\begin{align}
    \left(\frac{\dot{a}}{a}\right)^2&=\frac{8\pi\rho}{3}+\frac{\Lambda}{3}-\frac{k}{a^2} \label{eq:friedmann1} \,, \\
    \frac{\ddot{a}}{a}&=-\frac{4\pi}{3}\left(\rho+3p\right)+\frac{\Lambda}{3}
    \label{eq:friedmann2} \,,
\end{align}
where $\Lambda$ is the cosmological constant and a dot over a variable denotes the derivative with respect to cosmological time $t$.

Equations (\ref{eq:friedmann1}) and (\ref{eq:friedmann2}) are the so-called \emph{Friedmann equations} and imply the energy-momentum conservation
\begin{equation}
    \dot{\rho}+3\frac{\dot{a}}{a}(\rho+p)=0 \,,
    \label{eq:conservation}
\end{equation}
which is satisfied by all matter. Since the conservation equation is implied by the first two equations, one may notice that we have two equations but we seek three functions, namely $a(t)$, $\rho(t)$ and $p(t)$. Hence, our system is under-determined. To close the system, we impose a linear relation between $\rho$ and $p$ (the so-called {\it equation of state}), that is,
\begin{equation}
    p=w\rho \,, 
    \label{eq:EOS}
\end{equation}
where $w$ is a constant called the \emph{equation of state parameter}. Note that this is the equation of a \emph{baryotropic fluid}, i.e.~a fluid whose density depends only on pressure and not on other quantities (e.g.~the entropy). A few commonly considered values for the equation of state parameter are~\cite{Nemiroff:2007xs}: 
\begin{itemize}
\item $w=-1$: cosmological constant or dark energy, where the cosmological constant appearing in~(\ref{eq:friedmann1}) and~(\ref{eq:friedmann2}) can be interpreted as a perfect fluid with $p=-\Lambda/(8\pi)=-\rho$;
\item $-1<w<-1/3$: dynamical dark energy models;
\item $w=-1/3$: negative spatial curvature, where the curvature term appearing in (\ref{eq:friedmann1})
can be interpreted as a perfect fluid with $p=-k/(8\pi a^2)=-\rho/3$;
\item $w=0$: matter, sometimes referred to as ``dust'';
\item $w=1/3$: radiation;
\item $w=1$: stiff matter, for which the speed of sound equals the speed of light.
\end{itemize}

Combining \eqref{eq:conservation} with \eqref{eq:EOS} yields
\begin{align}
     \frac{\dot{\rho}}{\rho}=-3(1+w)\frac{\dot{a}}{a} &\implies \dv{t}\left(\log\rho\right)=-3(1+w)\dv{t}\left(\log{a}\right)
     \nonumber \\ &
     \implies \log{\rho}=-3(1+w)\log{a}+C \,,
\end{align}
where $C$ is a constant of integration. This means we find the relation
\begin{equation}
  \rho=\rho_0 a^{-3(1+w)} \,, 
  \label{eq:rhoanda}
\end{equation}
where $\rho_0$ is a constant of integration. In the case of a vanishing cosmological constant ($\Lambda=0$), we can combine equation \eqref{eq:friedmann1} with \eqref{eq:rhoanda}, to obtain
\begin{equation}
    \left(\frac{\dot{a}}{a}\right)^2 =
    \frac{8\pi\rho_0}{3}a^{-3(1+w)}-\frac{k}{a^2} \,.
\end{equation}
Multiplying through by $a^2$ and setting, for simplicity, $\beta=8\pi\rho_0/3$ and $\gamma=1+3w$, one has
\begin{equation}
    \left(\dot{a}\right)^2=\beta a^{-\gamma}-k \,. 
    \label{eq:f1}
\end{equation}
Equations~(\ref{eq:rhoanda}) and~(\ref{eq:f1}) require restrictions on the equation of state parameter $w$, or equivalently $\gamma$. We require closed cosmological solutions which recollapse. This implies that $\gamma >0$ or $w >-1/3$, we will see this condition emerging below. Moreover, we require $w \leq 1$ to ensure that the speed of sound is bounded by the speed of light. The physically meaningful parameter range is thus $-1/3 < w \leq 1$. In the following we also set $\Lambda=0$ for simplicity but note that the inclusion of the cosmological constant would be a natural extension of this work. 

\section{Properties of the solution}

Firstly, we note that setting $\dot{a}=0$ we obtain the maximal value of $a(t)$, that is,
\begin{equation}
    \label{eqn:amax}
    a_{\mathrm{max}}=\left(\frac{\beta}{k}\right)^{1/\gamma}.
\end{equation}
Since $\beta=8\pi\rho_0/3>0$, the only meaningful value is indeed obtained when $k=1$ (i.e.\ in the case of a ``closed'' universe). This corresponds to some time $t_{\rm max}$. The value of this time depends on the chosen initial conditions and we will not need that value for what follows. To see that $a_{\rm max}$ is indeed a local maximum, we consider~(\ref{eq:friedmann2}) and evaluate $\ddot a$ at $t_{\rm max}$, which gives
\begin{align}
    \ddot{a}_{\rm max} &= -\frac{4\pi}{3}(1+3w)\rho_{\rm max} a_{\rm max} \\
    &= -\frac{4\pi}{3}\rho_0 (1+3w) (a_{\rm max})^{-3(1+w)} a_{\rm max} \\
    &= -\frac{4\pi}{3}(1+3w) \rho_0 (a_{\rm max})^{-2-3w} < 0
\end{align}
provided $(1+3w)>0$ or $w>-1/3$ which we already stated above. In the second step, we use the conservation equation~(\ref{eq:rhoanda}). Now that we have established the existence of a maximum value, we will show that the solution is symmetric relative to that maximum. To do so, we will show that the part of the solution for $t$ less than $t_{\rm max}$ satisfies the same differential equation as the part for $t$ greater than $t_{\rm max}$. Let us set
\begin{align}
    L(t) &= a\left(t_{\rm max} - t\right) \eqqcolon a\left(\tau_-\right),
    \quad 0 \leq t \leq t_{\rm max}\\
    R(t) &= a\left(t_{\rm max} + t\right) \eqqcolon a\left(\tau_+\right),\quad 0 \leq t
\end{align}
which satisfy $a_{\rm max}=a(t_{\rm max})=L(0)=R(0)$, this means they coincide at $t=0$. We can view $L(t)$ as the `left' part of the solution and $R(t)$ as its `right' part. Thus, we have $\dot{L}=-\dot{a}$, $\dot{R}=\dot{a}$, and $-\dd  t=\dd\tau_-$, $\dd t=\dd\tau_+$. Now, we go back to equation \eqref{eq:f1} and note
\begin{align}
    \left(\dv{a}{t}\right)^2 &= \beta a(t)^{-\gamma}-k \\
    \iff
    \left(\dv{a}{\tau_-}\right)^2 &= \beta a(\tau_-)^{-\gamma}-k \\
    \iff
    \left(-\dv{L}{t}\right)^2 &= \beta L(t)^{-\gamma}-k \\
    \iff 
    \left(\dv{L}{t}\right)^2 &= \beta L(t)^{-\gamma}-k 
\end{align}
and similarly for $R$. Hence, both the `left' part and the `right' part satisfy the same differential equation and take the same value at $t=0$. By uniqueness, we must therefore have $L=R$ everywhere. This proves that the solutions is symmetric with respect to $t_{\rm max}$. This also implies that $L(t_{\rm max}) = R(t_{\rm max}) = a(0)$. We now assume $a(0)=0$, which means the complete solution $a(t)$ vanishes for different times, namely at $t=0$ and at $t=2t_{\rm max}=:t_{\rm end}$. We note that the ODE~\eqref{eq:f1} becomes singular when $a \to 0$, however, this has no bearing on our results. This goes back to the presence of the square root and simply implies that uniqueness does not hold at $a \to 0$ as there are two viable roots.

From the point of view of cosmology, this means our cosmological model has the following generic behaviour: the scale factor $a(t)$ vanishes at $t=0$, then sharply increases ($\dot{a}\to\infty$ as $t \to 0$) up to a maximum value $a_{\rm max}$. Respectively, these correspond to the big bang and the subsequent expansion of the universe. After reaching $a_{\rm max}$, the universe collapses with its evolution ending in the big crunch, where $a \to 0$ as $t \to t_{\rm end}$. Since this model has a finite lifetime, it is meaningful to consider the total angular distance travelled by an azimuthal geodesic released at the beginning of the universe, or the Big Bang. 

\section{Azimuthal null geodesics}

Light-like particles move along null geodesics if not subject to any force. Azimuthal null geodesics are defined by $r=r_0$ and $\theta=\theta_0$, both being constants. The physical picture is that of a signal being emitted at the beginning of the universe travelling along the angular direction only. We set $\theta_0=\pi/2$ and $r_0=1$, so that the line element~\eqref{FLRW} yields
\begin{equation}
    0=-\dd t^2+a^2(t)\dd\varphi^2 \,,
\end{equation}
which we can solve for the angular variable and integrate. Let us assume that the azimuthal null geodesic is released at $t=0$. We are interested in the total angular distance $\Delta\varphi$ travelled during the evolution of the universe up to the big crunch when $t=t_{\rm end}$, which is given by
\begin{equation}
    \Delta\varphi= \int\limits_{\varphi(0)}^{\varphi(t_{\rm end})} \dd\varphi=
    \int\limits_{0}^{t_{\rm end}}\frac{\dd t}{a(t)} \,,
    \label{eq:angle}
\end{equation}
where $a(t)$ is the solution to~\eqref{eq:f1} which satisfies $a(0)=0$. It is convenient to use the symmetry of the solution about its maximum and write
\begin{equation}
    \Delta\varphi = 2\int\limits_{0}^{t_{\rm max}}\frac{\dd t}{a(t)} \,. \label{eq:anglewithtmax}
\end{equation}

Going back to equation \eqref{eq:f1}, for $0\leq t\leq t_{\rm max}$, we have that
\begin{equation}
   \dv{a}{t}= \sqrt{\beta a^{-\gamma}-1} \quad \implies \quad
   \dd t= \frac{\dd a}{\sqrt{\beta a^{-\gamma} -1}}\,.
\end{equation}
The negative root corresponds to the contraction phase $t_{\rm max}\leq t\leq t_{\rm end}$. Therefore we arrive at
\begin{equation}
    \Delta\varphi = 2\int\limits_{0}^{t_{\rm max}} \frac{\dd t}{a(t)} =
    2\int\limits_{0}^{a_{\rm max}} 
    \frac{\dd a}{a\sqrt{\beta a^{-\gamma} -1}} \,.
\end{equation}
We shall employ the substitution $y=\sqrt{\beta a^{-\gamma}-1}$ or $a^{-\gamma}=\frac{1}{\beta}\left(1+y^2\right)$. Thus
\begin{equation}
    \dv{y}{a}=-\frac{\beta  \gamma  a^{-\gamma -1}}{2 \sqrt{\beta  a^{-\gamma }-1}}
    \quad \implies \quad 
    \dd a =-\frac{2y}{\beta\gamma a^{-\gamma -1}}\dd y \,.
\end{equation}
This reduces the integral to a standard integration problem 
\begin{align}
    \int \frac{\dd a}{a\sqrt{\beta a^{-\gamma} -1}} &= 
    -\frac{2}{\gamma}\int \frac{1}{1+y^2}\dd y \nonumber\\ &=
    -\frac{2}{\gamma}\arctan(y)+C =
    -\frac{2}{\gamma}\arctan\left(\sqrt{\beta a^{-\gamma}-1}\right)+C \,,
\end{align}
with some constant of integration $C$. It is now straightforward to evaluate the angular distance
\begin{equation}
    \Delta\varphi = 
    2\int\limits_{0}^{a_{\rm max}} \frac{\dd a}{a\sqrt{\beta a^{-\gamma} -1}} = 
    -\frac{4}{\gamma}\left[\arctan(\sqrt{\beta a^{-\gamma}-1})\right]_{0}^{a_{\rm max}}\,.
\end{equation}
As $a\rightarrow a_{\rm max}=\beta^{1/\gamma}$, due to~\eqref{eqn:amax}, we find $\arctan(\sqrt{\beta a^{-\gamma}-1})\rightarrow 0$. On the other hand, the lower limit, $a\rightarrow 0$, gives $\arctan(\sqrt{\beta a^{-\gamma}-1})\rightarrow \pi/2$. Recalling $\gamma=1+3w$, we therefore obtain the result
 \begin{equation}
     \Delta\varphi =\frac{2\pi}{1+3w}\,,
 \end{equation}
 valid for all $w>-1/3$. Interestingly, this angle can also be computed in the unphysical region where $w>1$. For such cosmological fluids the speed of sound would exceed the speed of light. However, these unphysical solutions eventually recollapse, which means that the angle can be computed.

\section{Conclusions}

The exact solution to the Einstein field equations for $a(t)$ in the case of a vanishing cosmological constant, and for arbitrary $w$, was first derived in~\cite{Assad:1986yh}, in terms of the hypergeometric function. Subsequently, it was shown in~\cite{Faraoni:1999qu} that the cosmological field equations can be rewritten as a Riccati differential equation, and this work was further extended by~\cite{Lima:2001fi}. For a non-vanishing cosmological constant,~\cite{Sonego:2011rb} proposed a qualitative study by considering ``effective potentials'', a common technique generally used in classical mechanics, and some exact solutions were presented and reviewed in the context of multi-fluid cosmology~\cite{Faraoni:2021opj}. \emph{A priori}, it might appear necessary to use explicit solutions for $a(t)$ and $t_{\rm max}$ to evaluate the azimuthal angle as defined in \eqref{eq:angle}. However, we have shown that the angular distance travelled by azimuthal geodesics can be computed explicitly for all $w$ without any knowledge about the explicit form of the solution. We extracted this result using only the properties of the ODE. 

A natural extension to this work would be the inclusion of a non-vanishing cosmological constant $\Lambda$. The value of $\Lambda$ would have to be chosen such that the solution eventually recollapses. As in the case without cosmological term, explicit solutions would involve special functions such as the Jacobi elliptic functions or the Weierstrass $\wp$-function. It is not clear in general whether the angular distance can be obtained as a simple formula as reported here, circumventing using explicit solutions.

It appears that very little is know about azimuthal geodesics in cosmological models.
Partially this can be explained by the fact that most standard cosmology is based on radial null geodesics as these can be related to observations. Nonetheless, the study of azimuthal geodesics seems interesting in its own right, especially from a more mathematical point of view. For the closed models we studied here, all azimuthal geodesics are great circles and, due to homogeneity and isotropy, all angular geodesics are in fact great circles. When considering the spatially flat or open cases, one will find different results as the topology will be no longer that of a 3-sphere. 

Often considered the ``Standard Model of Cosmology'', the $\Lambda$CDM model describes a spatially flat, cold dark matter (CDM) universe dominated by a positive cosmological constant. Recent research has shown some discrepancies between different observational data sets as well as with the $\Lambda$CDM model (see e.g.~\cite{DiValentino:2019qzk, Handley:2019tkm, Vagnozzi:2020rcz, Anselmi:2022uvj, Glanville:2022xes, Semenaite:2022unt, DiValentino:2020hov}). It may be interesting, therefore, to explore geodesic motion in FLRW models with non-standard choices of the parameters $k$, $\Lambda$ and $w$. 

\subsection*{Acknowledgements}

CGB wishes to acknowledge the contribution of Iona Duncan who worked on this topic with me for her MSci project \textit{Azimuthal Geodesics in Cosmology} in the academic year 2018/2019. We acknowledge fruitful discussions with Elias L.~M\"unch. Antonio d'Alfonso del Sordo is supported by the Engineering and Physical Sciences Research Council EP/R513143/1 \& EP/T517793/1.

\end{document}